\documentclass[twocolumn,preprintnumbers,amsmath,amssymb,aps,pra,longbibliography]{revtex4-1}
\usepackage{amsmath,amssymb,graphicx,epstopdf,bm,soul,upgreek, natbib}
\usepackage{graphicx}
\usepackage{color}
\definecolor{OliveGreen}{rgb}{0,0.6,0}
\usepackage{dcolumn}
\usepackage{bm}
\usepackage{amssymb,float}
\usepackage{cancel}
\usepackage{ulem}
\usepackage{siunitx}
\usepackage[colorlinks,bookmarks=false,citecolor=blue,linkcolor=blue,urlcolor=blue]{hyperref}
\begin{document}
\title{Topology with memory in nonlinear driven-dissipative photonic lattices}
\author{Subhaskar Mandal}\email[Corresponding author:~]{subhaskar.mandal@ntu.edu.sg}
\author{Gui-Geng Liu}
\author{Baile Zhang}\email[Corresponding author:~]{blzhang@ntu.edu.sg}

\affiliation{Division of Physics and Applied Physics, School of Physical and Mathematical Sciences, Nanyang Technological University, Singapore 637371, Singapore}

\begin{abstract}
We  consider a photonic lattice of nonlinear lossy resonators subjected to a coherent drive, where the system remembers its topological phase. Initially, the system is  topologically trivial. After the application of an additional coherent pulse, the intensity is increased, which modifies the couplings in the system and then induce a  topological phase transition. However, when the effect of the pulse dies out, the system does not go back to the trivial phase. Instead, it remembers the topological phase and maintains its topology acquired during the pulse application. The pulse can be used as a switch to trigger amplification of the topological modes. We further show that the amplification takes place at a different frequency as well as at a different position from those of the pulse, indicating frequency conversion and intensity transfer. Our work can be useful in triggering the different functionalities of  active topological photonic devices.
\end{abstract}

\maketitle

\section*{Introduction}
The intriguing properties of topological photonics  have enabled widespread applications in  modern optical devices, such as robust signal transport \cite{Wang2009,PhysRevLett.106.093903,Shalaev2019,He2019,Yang2020}, optical delay line \cite{Hafezi2011}, quantum interface \cite{doi:10.1126/science.aaq0327},  quantum light source \cite{Mittal2018}, robust splitters \cite{Cheng2016}, topological lasers \cite{St-Jean2017,doi:10.1126/science.aao4551,doi:10.1126/science.aar4005,Zeng2020,Smirnova2020}, etc. Topological photonics is also promising for optical information processing technologies. For example, valley  photonic crystals are identified as an excellent candidate for robust information transfer in next-generation devices \cite{Shalaev2019,He2019,Yang2020}. Similar to transferring the information, the ability to store them in  memory, is an equally important task in information processing. However, optical memories along with topological protection have not been explored till now.

Nonlinearity is at the core of memory devices. Interplay between the nonlinearity and the topology has made way for many novel effects such as topological solitons
 \cite{PhysRevLett.117.143901,Kartashov:16,doi:10.1126/science.aba8725,PhysRevX.11.041057,Pernet2022}, high harmonic generation \cite{Kruk2019,Wang2019}, topological phase transitions \cite{PhysRevB.93.020502, Hadad2018,Zhou_2017,PhysRevB.99.115423, PhysRevLett.123.053902,doi:10.1126/science.abd2033}, and others \cite{PhysRevLett.111.243905,PhysRevLett.119.253904,PhysRevLett.124.063901,doi:10.1126/science.abf6873,Kirsch2021,PhysRevB.104.235420} (see Ref.~\cite{doi:10.1063/1.5142397} for a comprehensive review). However, none of the previous works can show the memory feature: once the key ingredient, that induces the functionalities, is removed from the scheme, the systems can no longer continue to exhibit such effects.

In this work, we introduce for the first time a topological phase with memory in a lattice of lossy resonators having local onsite Kerr-nonlinearity, where the system remembers its  topological phase.  The lossy nature of the system leads to  a steady-state in the presence of a coherent drive $F$. However, due to the nonlinearity, our system subjected to a properly designed $F$ shows not only one but two steady states: low and high-intensity states. Our system is topologically trivial and after $F$ is introduced, it attains the low-intensity steady-state. The introduction of an additional coherent pulse increases the intensity of the system. At higher intensities, the nonlinear interaction modifies the couplings and the otherwise-trivial system becomes topological.  However, at longer times when the effect of the pulse dies, the system does not go back to its previous trivial phase. Instead, it remembers the topological phase and maintains its topology acquired during the pulse application. As an application of this effect, we show a unique amplification phenomenon, where the amplification is triggered by a pulse.

\begin{figure}[t]
\includegraphics[width=0.5\textwidth]{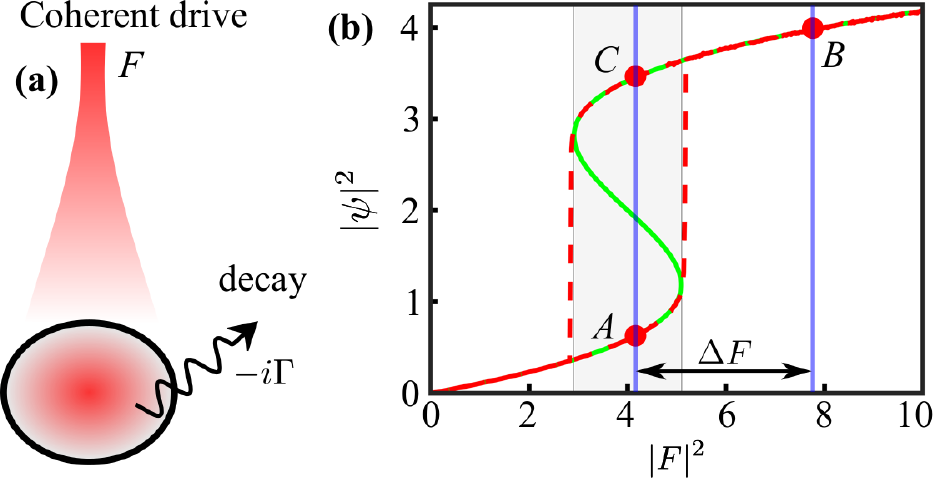}
\caption{(a) Schematic of a coherently driven nonlinear resonator. (b) Analytically and numerically calculated bistable curves in green and red, respectively. Parameters: $\Delta=-3$, $\Gamma=1$.}
\label{Fig1}
\end{figure}

We start by considering a nonlinear optical resonator subjected to a coherent drive $F$ (see Figure~\ref{Fig1}a), which is represented by the following nonlinear schr\"odinger equation (NLSE)
\begin{align}\label{Eq1}
i\frac{\partial \psi}{\partial t}=\left(\omega_0-i\Gamma\right)\psi+|\psi|^2\psi+F\exp(-i\omega_p t).
\end{align}
Here $\omega_0$ is the onsite potential and $\Gamma$ is the linear decay. The next term represents the defocusing Kerr nonlinearity, where the nonlinear coefficient is set to 1. $F$ is a coherent drive having frequency $\omega_p$. For the steady-state $\psi_s$, where $\partial \psi_s/\partial t=0$, one can obtain
\begin{align}\label{Eq2}
\left|F\right|^2=\left[\left(\Delta+\left|\psi_s\right|^2\right)^2+\Gamma^2\right]\left|\psi_s\right|^2,
\end{align}
where $\Delta=\left(\omega_0-\omega_p\right)$. From eq.~(\ref{Eq2}), it is easy to find that within the gray region for each $|F|^2$ three possible $|\psi_s|^2$ exist (see Figure~\ref{Fig1}b). However, in practice, the middle branch is not stable. This can be confirmed by numerically solving eq.~(\ref{Eq1}), but letting $F$ vary very slowly in time such that at each time step steady state can be reached (more details on the numerical calculation of bistability can be found in the supporting information). The red dashed curve obtained numerically follows the analytical green curve, however, the middle branch does not appear. Consequently, the system shows bistability by allowing both the low and high-intensity stable states for a fixed value of $F$ within the gray region.  The absence of the middle branch of the Bistability curve can be explained using the stability analysis based on the first Lyapunov Criterion \cite{https://doi.org/10.1002/advs.201900771} (see supporting information).

An important characteristic of bistability is their ability to mimic the memory: the state of the system is not only determined by the current parameters (such as $F$) but also by its previous state. For example, let us consider a system that is initially in the low-intensity state ``$A$" as shown in Figure~\ref{Fig1}b. An additional coherent drive $\Delta F$ is added such that the system moves to a high-intensity state ``$B$". Now if $\Delta F$ is removed, the system does not return back to its original state ``$A$", instead, it chooses the high-intensity state ``$C$". While determining the final state, the system memorizes the information (high-intensity) about the intermediate state ``$B$" and in the case where ``$B$" is a low-intensity state, the system would return to the original state ``$A$" upon removing $\Delta F$. 

\section*{Model}
\begin{figure}[t]
\includegraphics[width=0.5\textwidth]{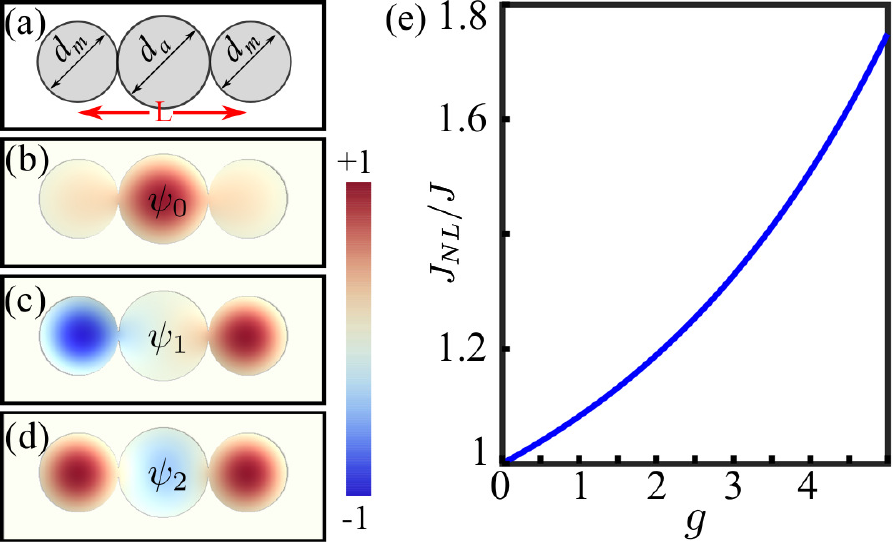}
\caption{(a) Schematic of two {\it main} resonators connected by an {\it auxiliary} resonator. (b,c,d) The lowest three modes of the system shown in (a). (e) The ratio of the coupling in the nonlinear regime $J_{NL}$ to the linear coupling $J$ between two {\it main} resonators as a function of the intensity of the {\it auxiliary} resonator.   Parameters: $V_0=236$, $L=2.13$, $d_m=1$, $d_a=1.13$.}
\label{Fig2-1}
\end{figure}

We arrange the nonlinear resonators in a 2D lattice. In order to be close with experiments, we model the dynamics of the system using the NLSE in the continuum limit (where the space is taken as continuous). Without the loss of generality, we work with the dimensionless NLSE, which is expressed as 
\begin{align}\label{Eq3}
 i\frac{\partial \psi(x,y)}{\partial t}=&\left[-\nabla^2+V(x,y)-i\Gamma\right]\psi(x,y)\notag\\
 &+\left|\psi(x,y)\right|^2\psi(x,y)+F(x,y)e^{-i\omega_pt}\nonumber\\
 &+F_p(x,y)\exp\left[-\frac{(t-t_0)^2}{2\tau^2}\right]e^{-i\omega_pt}. 
 \end{align}
Here $\nabla^2\equiv \left( {\partial^2}/{\partial x^2}+{\partial^2}/{\partial y^2}\right)$ is the transverse Laplacian operator, $V$ is the external potential profile corresponding to the resonators, $\Gamma$ is the linear decay, $F$ is a position-dependent coherent pump having frequency $\omega_p$, and  $F_p$ is a coherent pulse having duration $\tau$ centered at time $t_0$. Next, we consider circular resonators having diameter $d_m$, which we call  {\it main} resonators. Two {\it main} resonators separated by $L$ are coupled via an {\it auxiliary} larger resonator having diameter $d_a$, where $d_a>d_m$ (see Figure~\ref{Fig2-1}a).  The potential is taken as $V=0$ inside and $V=V_0>0$ outside the resonators.  

\begin{figure*}[t]
\includegraphics[width=0.7\textwidth]{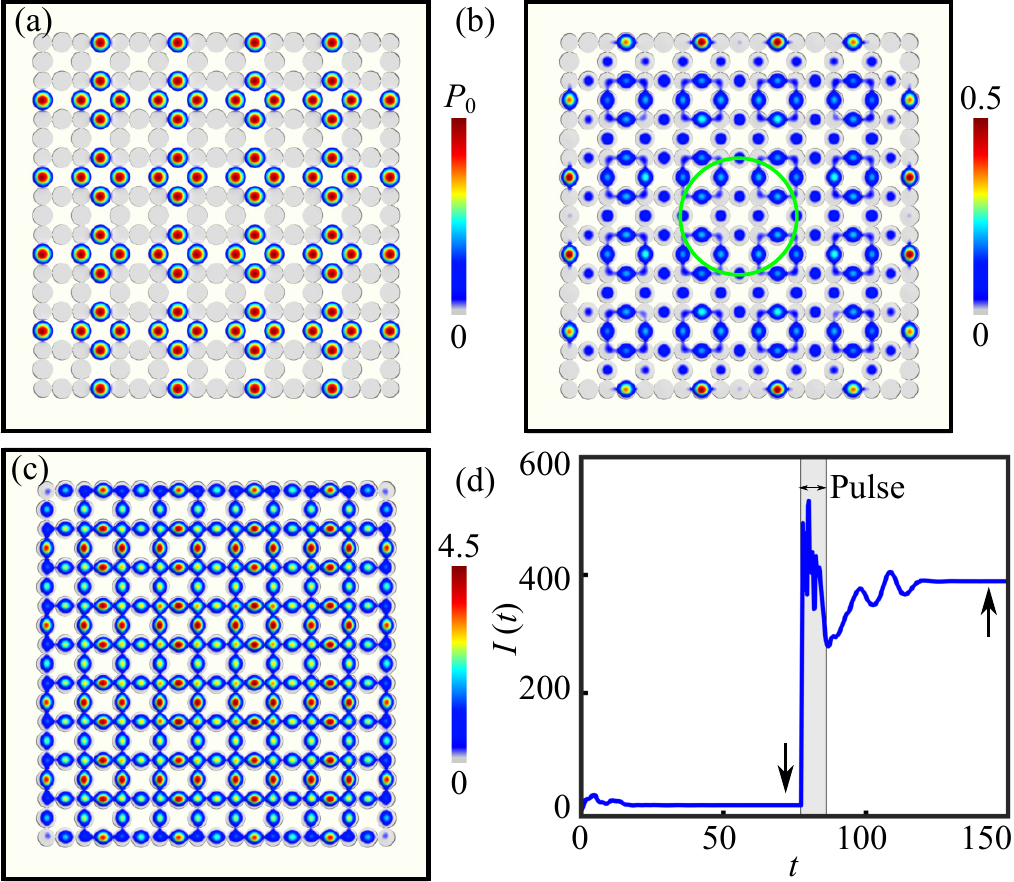}
\caption{(a) Spatial profile of the coherent pump. (b,c) Steady-states of the system before and after the coherent pulse $F_p$, respectively. The green circle in (b) represents the width of $F_p$. (d) The total intensity of the system as a function of time showing the bistable behavior. The two arrows indicate the times at which the states in (b,c) are plotted.
Parameters: $\Gamma=0.13$, $P_0=\sqrt{0.5}$, $\omega_p=14.74$, $\sigma=0.3$, $\tau=23.9$, $t_0=77$. $F_p$ is a gaussian pulse having strength $20P_0$ and width $7.5\sigma$. All other paramters are kept the same as those in Figure~\ref{Fig2-1}.}
\label{Fig2}
\end{figure*}

To capture the role of nonlinearity, we first consider two {\it main} resonators connected by an {\it auxiliary} resonator as shown in Figure~\ref{Fig2-1}a. The ground state wave function $\psi_0$ is mainly localized at the {\it auxiliary} resonator (see Figure~\ref{Fig2-1}b), whereas the first ($\psi_1$) and second ($\psi_2$) excited states are  localized at the {\it main} resonators (see Figure~\ref{Fig2-1}c,d). The coupling strength $J$ between the {\it main}  resonators can be estimated from the difference in the eigenvalues of the symmetric ($E_2$) and asymmetric ($E_1$) eigenstates, where $J=\left(E_2-E_1\right)/2$. The important feature that plays a key role and signifies the nonlinear effect  in this work is the ability to control the coupling between the {\it main} resonators by changing the intensity of the {\it auxiliary} resonator. This is captured by choosing an effective potential $V_{\text{eff}}=V+g\left|\psi_0\right|^2$, where $g$ corresponds to the peak value of the ground state intensity, and obtaining the coupling in the nonlinear regime $J_{NL}$ in a self-consistent way. Figure~\ref{Fig2-1}e shows the enhancement of $J_{NL}$ compared to $J$ as a function of $g$. However, such an enhancement of $J_{NL}$ is limited. Once $g$ becomes larger than the difference between the fundamental frequencies of the {\it main} and  the {\it auxiliary} resonators, $J_{NL}$ would start to decrease and for $g\rightarrow\infty$, $J_{NL}\rightarrow0$.

\begin{figure*}[t]
\includegraphics[width=0.7\textwidth]{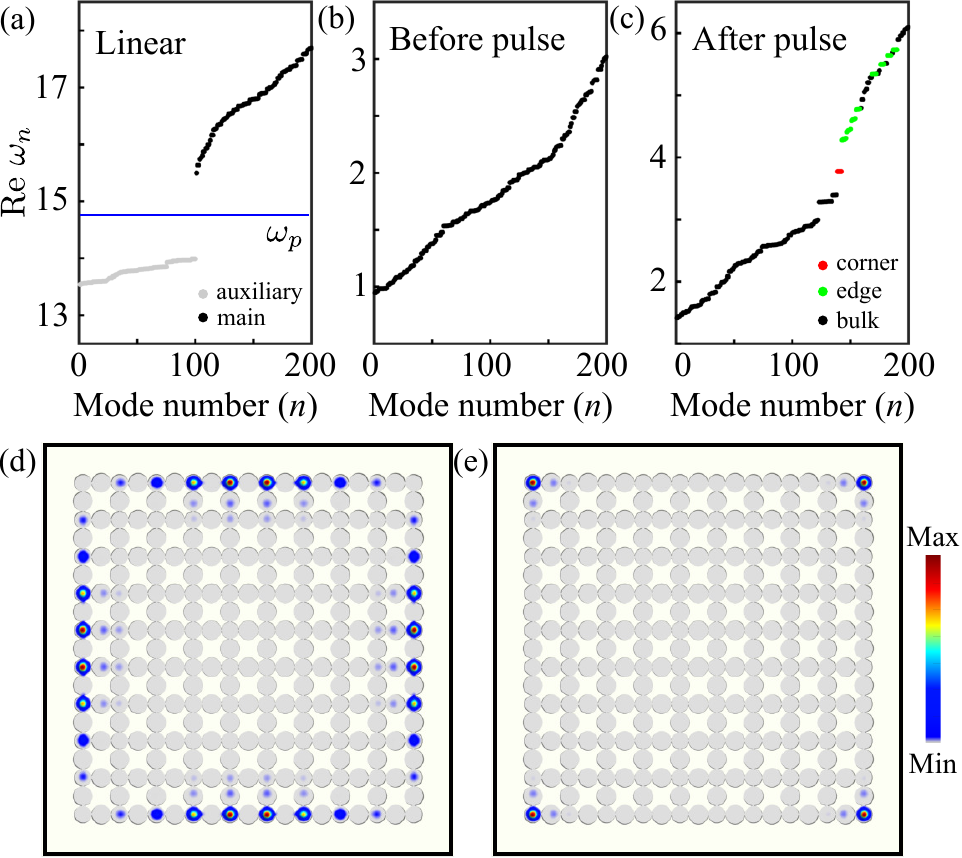}
\caption{(a,b,c) Real eigenfrequencies for different cases. In (a) the {\it auxiliary} resonator band is shown in gray and the {\it main} resonator band is shown in black. Red and green dots in (c) correspond to the corner and edge modes, respectively. The blue line in (a) represents $\omega_p$, with respect to which (b,c) are rescaled. (d,e) Spatial profiles of an edge and a corner mode, respectively. All the parameters are kept the same as those in Figure~\ref{Fig2}.}
\label{Fig3}
\end{figure*}

\section*{Topological memory}
Now that we have all the ingredients, we proceed to study the topological phase in a 2D square lattice formed by the above-mentioned resonators, where between any two {\it main} resonators there is an {\it auxiliary} resonator. Recalling that the intensity of the  {\it auxiliary} resonators enhances the coupling between the {\it main} resonators, we choose the coherent pump profile in such a way that the inter-cells are coupled strongly similar to the 2D Su-Schrieffer–Heeger (SSH) model \cite{PhysRevB.98.205147}. The spatial profile of the coherent  pump is expressed as
\begin{align}\label{Eq4}
F(x,y)=P_0\sum_{X_n,Y_n}\exp \left[-\frac{(x-X_n)^2+(y-Y_n)^2}{2\sigma^2}\right],
\end{align}
where $P_0$ is the strength of the pump and $\{X_n,Y_n\}$ are the coordinates of the center of the pumped {\it auxiliary} resonators as shown in Figure~\ref{Fig2}a. Such pump profiles are readily achieved in practice using the spatial light modulators \cite{Moreno:12,Topfer:21}. We choose the value of $P_0$ such that the {\it auxiliary} resonators subjected to the pump are bistable. To show the bistable behaviour of the whole system, we solve eq.~(\ref{Eq3}) without the pulse ($F_p=0$) and take zeros as the initial condition. While plotting the spatial profiles, we keep the background potential to distinguish the intensities between the {\it main} and {\it auxiliary} resonators. Figure~\ref{Fig2}b shows the steady-state of the system before the application of the pulse, where intensity is mainly localized at the pumped {\it auxiliary} resonators.  The coupling between the resonators results in slightly non-identical bistability curves of the resonators placed at different positions (see supporting information).  Due to this, the intensity among the pumped resonators varies a little, but they remain in the low-intensity state, where $|\psi|^2$ is negligible.

Next, we apply a gaussian-shaped coherent pulse $F_p$ centered at the bulk as shown by the green circle in Figure~\ref{Fig2}b. The addition and removal of the additional pump $\Delta F$ in Figure~\ref{Fig1}b is played by the pulse $F_p$ here. In Figure~\ref{Fig2}c the steady-state of the system after the application of the pulse is shown. The system indeed remembers the high intensity created by the pulse and once the effect of the pulse dies out the system chooses to stay at the high-intensity state. Compared to the low-intensity state in Figure~\ref {Fig2}b, a much larger intensity outside the pumped {\it auxiliary} resonators exists, which signifies the enhancement of the coupling due to significant $|\psi|^2$. In Figure~\ref {Fig2}d the total intensity of the system, $I(t)=\int\psi(x,y,t)~dxdy$, where the integration is over the whole system, is shown as a function of time, which shows the bistable behaviour of the system. The full dynamics of the system is shown in movie1.

We have performed all the calculations corresponding to the 2D lattice on a $2^9\times2^9$ grid. The Laplacian is taken into account through the FFT (Fast Fourier Transform) spectral method. It should be noted that the finite difference (FD) method can also be implemented to express the Laplacian. However, FD requires larger computational memory and is more time-consuming compared to the FFT method. The time dynamics is performed using  Matlab’s ODE solvers, which relies on well-established numerical techniques, such as the explicit Runge-Kutta (4,5) formula, the Dormand-Prince pair 
\cite{DORMAND198019}.

Having established the memory effect in the 2D lattice, here we show the topology associated with it. In Figure~\ref{Fig3}a the eigenfrequencies of the linear system are shown, which can be found by putting the nonlinear and pumping terms to zero in eq.~(\ref{Eq3}) and diagonalizing its corresponding Hamiltonian. The lower band (shown in gray) has the main contributions from to the {\it auxiliary} resonators, whereas the upper band has the main contributions from the {\it main} resonators.  For the rest of the work, we shall focus on the {\it main} resonator band, which is topologically trivial and gapless in the linear regime.  To include the nonlinear effect, we study the Bogoliubov fluctuations on top of the steady-state \cite{Pitaevskii_Stringari_2003}:
\begin{align}\label{Eq5}
\psi(x,y)=\psi_s(x,y)+u_n(x,y)e^{-i\omega_n t}+v_n^*(x,y)e^{i\omega_n^* t}.
\end{align}
Here $\psi_s$ represent the low and high-intensity steady states shown in Figure~\ref{Fig2}b,c, respectively. $u_n$ and $v_n$ represent the fluctuations having frequency $\omega_n$. Substituting eq.~(\ref{Eq5}) into eq.~(\ref{Eq3}) and by ignoring the higher-order terms in $u_n$ and $v_n$, we obtain the following eigenvalue equation
\begin{equation}\label{Eq6}
\begin{bmatrix}
H_0+2\left|\psi_s\right|^2 & \psi_s^2\\
-{\psi_s^2}^* & -H_0^*-2\left|\psi_s\right|^2
\end{bmatrix}
\begin{bmatrix}
u_n\\
v_n
\end{bmatrix}=
\omega_n
\begin{bmatrix}
u_n\\
v_n
\end{bmatrix},
\end{equation}
where $H_0=\left[-\nabla^2+V(x,y)-i\Gamma-\omega_p\right]$, which is rescaled with respect to the pump frequency $\omega_p$.

The fluctuation Hamiltonian  has particle-hole symmetry, which makes the eigenfrequencies appear in pairs ($\omega_n,~-\omega_n^*$). The Hilbert space of the fluctuation Hamiltonian is double in size compared to the linear one. Consequently, for better visualization we show the eigenfrequencies near the {\it main} resonator band and for Re $\omega_n>0$. Figure~\ref{Fig3}b shows the eigenfrequencies of the fluctuations before the pulse is applied. Similar to the linear case, the band is gapless. This is understandable as for the low-intensity state, $|\psi|^2$ is not significant enough to induce the topological transition. After the application of the pulse positioned at the bulk, the system switches to the high-intensity state. In this case, the nonlinear effect becomes significant and the coupling between the {\it main} resonators connected by a pumped {\it auxiliary} resonator increases. The system goes through a topological phase transition from trivial to second-order topological phase, where a bulk band gap opens and four topological corner modes appear. An effective tight-binding model based on the strong coupling induced by the pump can reproduce the topological corner modes (see supporting information).

In Figure~\ref{Fig3}c the eigenfrequencies of the fluctuation after the application of the pulse are shown, where the topological corner modes are marked in red and the edge modes are shown in green. Figure~\ref{Fig3}d,e show the spatial profiles of one of the topological edge modes ($n=146$) and topological corner modes ($n=142$), respectively. In experiment, the topological modes will be hidden in the high-intensity steady state. However, they can be probed using a weak additional coherent pump followed by the frequency filtration to subtract the steady-state.

Higher order topological phases have been an intense area of research \cite{doi:10.1126/science.aah6442,PhysRevLett.122.076801,PhysRevB.98.205147}. We note that the effect of the on-site Kerr-nonlinearity on these system have been studied recently \cite{PhysRevLett.124.063901,PhysRevB.104.235420,Hu2021,Kirsch2021}. However, the previous works consider the effect of nonlinearity on the already existing linear topological band structure. The presented result is the first example, where onsite Kerr-nonlinearity alone induces higher-order topological phase transition. Although here we have focused on the topological corner modes, it would be interesting to investigate further whether different truncations can lead to different types of edge states, such as the ones shown in Ref.~\cite{PhysRevB.105.L201105}.
   
\section*{Calculation of the topological invariant}

\begin{figure}[t]
\includegraphics[width=0.45\textwidth]{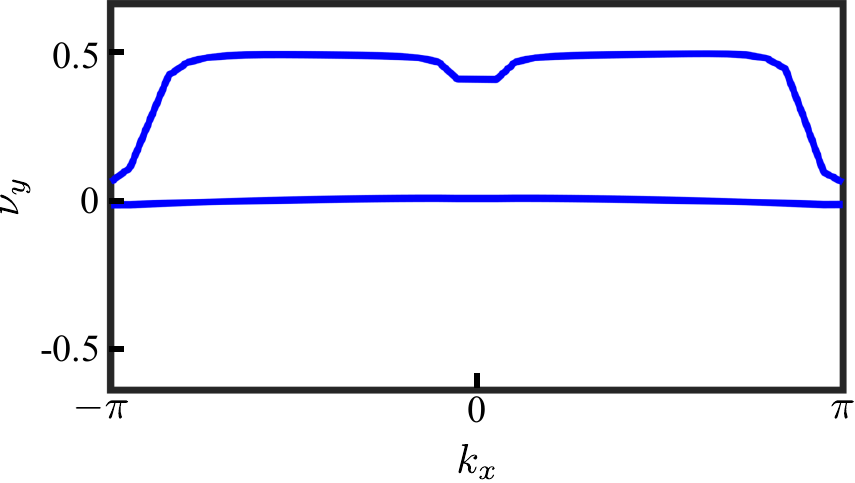}
\caption{Plot of the Wannier bands obtained by diagonalizing eq.~(\ref{AEq9}).}
\label{AFigWilsonLoop}
\end{figure}

Here we calculate the bulk polarization, which characterizes the topological corner modes \cite{PhysRevB.96.245115} using a bi-orthogonal Wilson loop \cite{PhysRevLett.123.073601}. We define a Wilson line along the $y$-direction as
\begin{align}
\Xi^{m,n}_y(k_x,k_y)&=\langle\Psi^{L,m}_{k_x,k_y+\Delta_y}|\Psi^{R,n}_{k_x,k_y}\rangle\notag\\
&=\int_{\text{unit cell}}\Psi^{L,m*}_{k_x,k_y+\Delta_y}(x,y)\Psi^{R,n}_{k_x,k_y}(x,y)~dxdy,
\end{align}
where $\Psi^{n}_{k_x,k_y}$ represents the Bloch eigenvectors of the fluctuations (see supporting information for details on the Bloch eigenvector calculation), $L$ and $R$ correspond to the left and right  Bloch eigenstates, respectively. $\Delta_y=2\pi/N_y$, where $N_y$ is the total number of points used along the $y$ direction of the Brillouin zone for calculation. Taking the periodicity as one, the Wilson loop along the $y$ direction is defined as
\begin{align}
W^y(k_x,k_y)=&\Xi^{m,n}_y(k_x,k_y+2\pi).....\Xi^{m,n}_y(k_x,k_y+2\Delta_y)\notag\\
&\Xi^{m,n}_y(k_x,k_y+\Delta_y)~\Xi^{m,n}_y(k_x,k_y).
\end{align}
The Wannier Hamiltonian is given by
\begin{align}\label{AEq9}
H_W^y(k_x,k_y)=-\frac{i}{2\pi}\log\left[W^y(k_x,k_y)\right].
\end{align}
The eigenvalues of $H_W^y$ form the Wannier bands, which are shown in Figure~\ref{AFigWilsonLoop}. The bulk polarization  $P=(P_x,P_y)$ is the same as the Wannier center. For a topologically trivial system $P=(0,0)$. The $x$ component of the bulk polarization $P_x$ is given by summing all the eigenvalues $\nu_y$ corresponding to all the momentum $k_x$ \cite{PhysRevB.96.245115}. It can be seen from  Figure~\ref{AFigWilsonLoop} that $P_x\approx0.5$. Since the system has $C_4$ rotational symmetry, one can obtain a similar result by choosing the Wilson loop along the $x$-direction, such that the $y$ component of the bulk polarization $P_y\approx0.5$ and the total polarization becomes $P\approx(0.5,0.5)$ making the system topologically nontrivial.

\begin{figure*}[t]
\includegraphics[width=0.65\textwidth]{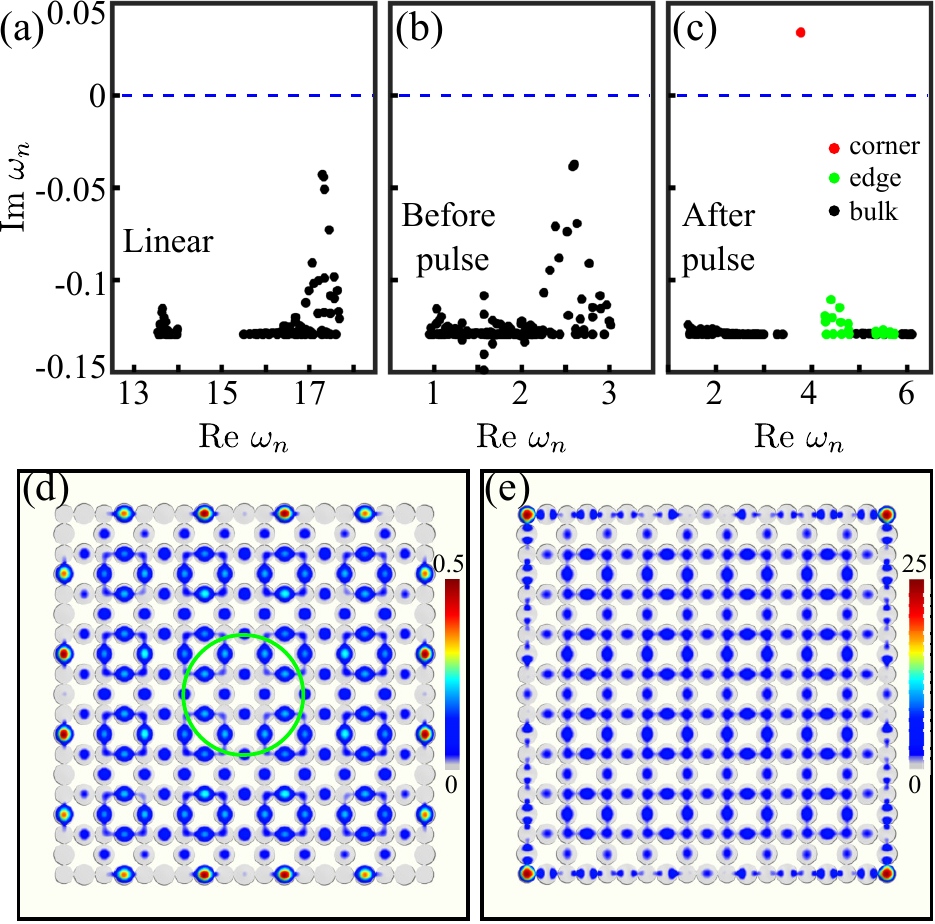}
\caption{(a,b,c) Complex eigenfrequencies for different cases.  Steady states from Figure~\ref{Fig2}b,c are used for obtaining (b,c). (d,e) Steady states of the system with gain at the corners before and after the pulse is applied, respectively. The green circle in (d) shows the position and width of the pulse. Parameters: Peak value of the gain $G_0=3\Gamma$. All other parameters are kept the same as those in the Figure~\ref{Fig2}.}
\label{Fig4}
\end{figure*}

\section*{Amplification of the corner modes}
As an application, we use our proposed scheme to control the functionality of an active topological photonic device. We introduce gain at the {\it main} four corner resonators of the 2D lattice. The gain is modeled by adding a term $+iG(x,y)\psi(x,y)$ at the right-hand side of eq.~(\ref{Eq3}), where $G(x,y)$ is composed of four Gaussians centered at the four corners having width $\sigma$ and peak value $G_0$. In this case, $H_0$ in eq.~(\ref{Eq6}) is updated to $H_0\rightarrow H_0+iG(x,y)$. To signify the role of nonlinearity, it is important that the gain alone can not induce lasing in the linear regime. Consequently, we obtain the complex eigenfrequencies for the linear system. The system stays below the lasing threshold Im~$\omega_n<0$ (see Figure~\ref{Fig4}a). Next, we take the steady states corresponding to  Figure~\ref{Fig2}b,c and obtain the same plot for the fluctuations before and after the pulse is applied. Similar to the linear case, the modes have Im $\omega_n<0$  before the application of the pulse (see Figure~\ref{Fig4}b). However, after the pulse is applied topological corner modes appear and due to the significant overlap with the gain, only they have Im $\omega_n>0$, while all other modes remains at Im $\omega_n<0$ (see Figure~\ref{Fig4}c). This has a significant effect on the steady-states. As predicted from the complex eigenfrequencies, the gain at the four corners does not alter the steady-state before the pulse, which is the same as the one obtained without the gain in Figure~\ref{Fig2}b. However, after the pulse is applied, a large intensity at the corners is observed along with the high-intensity steady state at the bulk (see movie2).

\begin{figure*}[t]
\includegraphics[width=\textwidth]{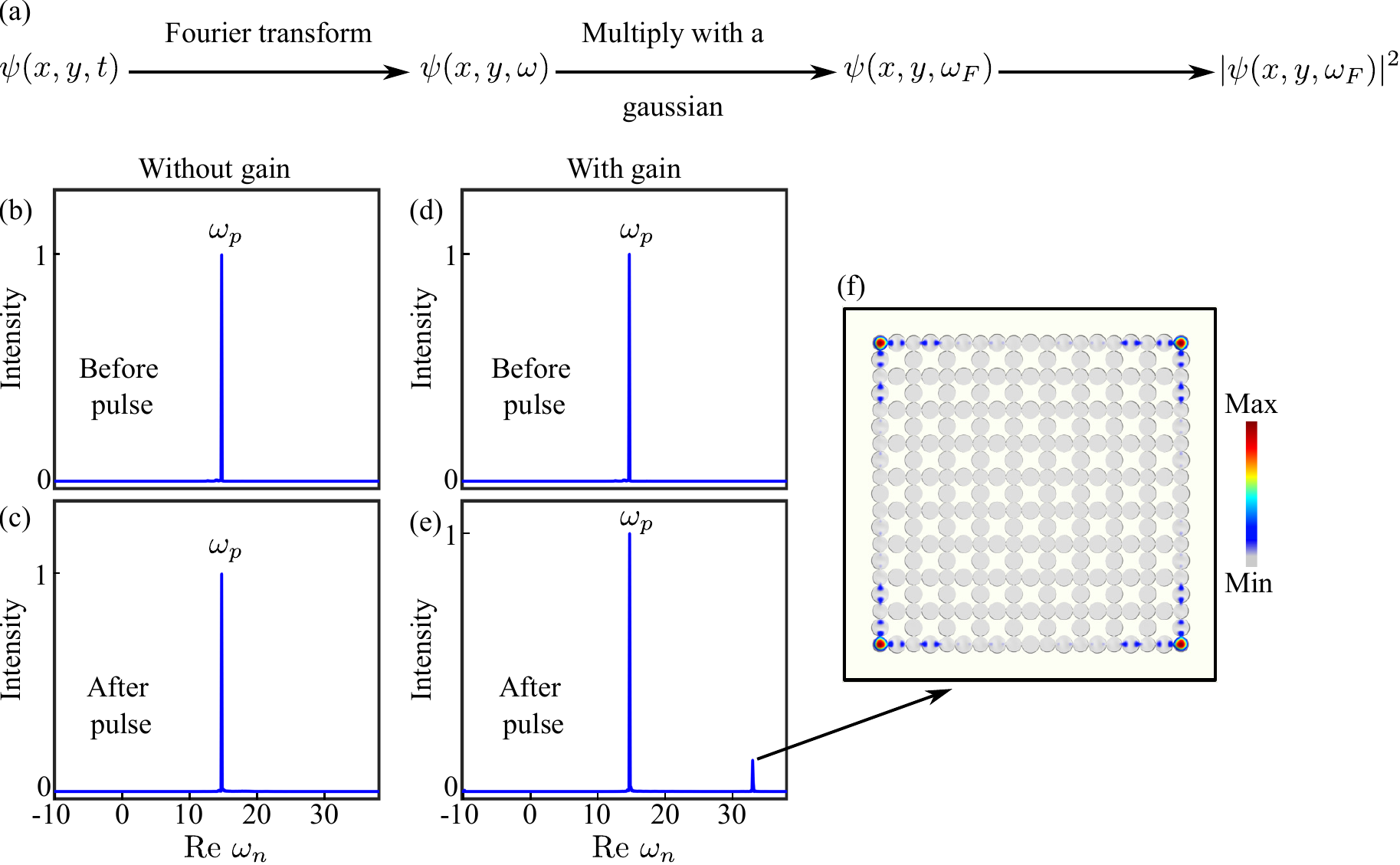}
\caption{(a) Steps to obtain spatial profiles of $\psi$ at a particular frequency by solving the NLSE. (b,c,d,e) Frequency-dependent intensity. The second peak in (e) corresponds to the amplification. (f) The spatial profile corresponds to the amplification peak in (e), which has the profile of corner modes. In each case we normalize the intensity with respect to the intensity at $\omega_p$.}
\label{Fig5_Filtered}
\end{figure*}

To confirm that the intensity at the corners correspond to the topological corner modes, we further obtain their spatial profile from the time-dynamics, which does not rely on the linear Bogoliubov theory (which does not include the higher-order terms in $u$ and $v$). We store the solutions at the intermediate time steps corresponding to the steady states in Figure~\ref{Fig4}d,e and then Fourier transform them along the time axis to move to the frequency dimension. At this stage, we can plot the intensity as a function of frequency. In order to obtain the spatial profile corresponding to a particular frequency, we first multiply with a gaussian to filter the desired frequency and plot the intensity. In Figure~\ref{Fig5_Filtered}a, the above-mentioned steps are given.  In Figure~\ref{Fig5_Filtered}b,c,d,e the frequency-dependent intensities are shown for different cases. Without the gain, we obtain a peak at the pumped frequency $\omega_p$ for both before and after the application of the pulse. It is important to note that no other peak is observed. Now we move on to the case where we introduce gain at the four corners. In this case also, before the application of the pulse a single peak at $\omega_p$ is obtained. However, after the pulse is applied a new peak along with the pumped peak is obtained (see Figure~\ref{Fig5_Filtered}e). It proves that the amplification at a different frequency has taken place after the application of the pulse. To prove that the amplified modes are the corner modes, we plot the spatial profile of the wave function around the amplified frequency by filtering out the desired frequency using the recipe mentioned in Figure~\ref{Fig5_Filtered}a. From the spatial profile shown in Figure~\ref{Fig5_Filtered}f, it is clear that the amplified modes are indeed the corner modes. Due to the presence of the $\alpha|\psi|^2$ term the corner modes are shifted in frequency from their linear Bogoliubov spectrum, which does not include the higher-order terms in $u$ and $v$.

There are a few unique feature associated with the amplification process presented above. Firstly, the pulse is positioned away from the corners at the bulk. However, the amplification takes place at the corners of the system, which do not overlaps with the pulse. This can be interpreted as the intensity transfer from the bulk to the corner. However, we stress that such a process is completely different from the signal transfer typically associated with the 1D edge modes of first-order topological insulators. Secondly, the amplified corner states have frequency different from that of the coherent drive, which makes this system suitable for frequency conversion. This is similar to the recent works on the high harmonic generation using topological systems \cite{Kruk2019,Wang2019,doi:10.1021/acsphotonics.1c01511,PhysRevLett.123.103901}, although here our underlying linear system is topologically trivial.  Lastly, the amplification, which is triggered by a pulse, is different from all the previous cases. For example, in lasers if the pumping term (or any other ingredient), which induces amplification (lasing) is removed, understandably the amplification would stop. Due to the memory effect, the functionality in our scheme remains, although the pulse disappears.

We note that a different type of topological memory is proposed recently in dynamic PT-symmetric optical resonators having saturable nonlinearity \cite{https://doi.org/10.1002/aelm.202200579,YuPiaoPark+2021+2883+2891}. Unlike the present case, a lattice is not involved there, instead, the system consists of a single diatomic or triatomic resonator. The topology introduced in those works is different from the topology associated with band structures in usual topological lattices. There oscillating quenching states are protected against the defects of the parameter space that preserve dynamical state trajectories.

\section*{Proposal for experimental realization}

The NLSE in eq.~(\ref{Eq3}) is generally used to describe the topological physics in photonic waveguide arrays \cite{doi:10.1126/science.aba8725,PhysRevX.11.041057}. Bistability is realized experimentally in waveguide arrays \cite{PhysRevLett.94.243902}, whereas the Bogoliubov fluctuations in eq.~(\ref{Eq6}) can be arranged using the parametric down-conversion \cite{PhysRevX.4.031007,Kruse_2013,Peano2016,PhysRevA.104.043710,PhysRevA.105.023513}. Alternatively, the system of exciton-polaritons, where cavity photons exhibit Kerr-nonlinearity by coupling strongly with the quantum well excitons, is also a promising platform for realizing our scheme. They are well known  for studying topological photonics \cite{PhysRevB.91.161413,PhysRevLett.114.116401,Klembt2018,PhysRevB.98.075412,PhysRevLett.122.083902,PhysRevLett.125.123902,doi:10.1021/acsphotonics.1c01425,doi:10.1021/acsphotonics.0c01958}.. Bistability is  well established for exciton-polaritons \cite{PhysRevA.69.023809,Paraiso2010,Banerjee_2020}. Bogoliubov fluctuations naturally arise in polariton system \cite{Utsunomiya2008,Pernet2022}. By choosing the proper physical units, our present parameters can be related to the exciton-polariton lattices.

Here we provide the physical parameters based on the exciton-polariton system.Typically, the system of exciton-polaritons is expressed using the NLSE (also known as the nonlinear Gross-Pitaevskii equation) \cite{RevModPhys.85.299}
\begin{align}\label{Eq18}
i\hbar \frac{\partial\phi(X,Y)}{\partial T}=& \left[-\frac{\hbar^2}{2m}\left(\frac{\partial^2}{\partial X^2}+\frac{\partial^2}{\partial Y^2}\right)\right.\notag\\
&\left.+V_e(X,Y)-i\frac{\hbar}{2\lambda}+\alpha|\phi(X,Y)|^2\right]\phi(X,Y)\nonumber\\
&+f(X,Y)\exp(-i\Omega_pt).
\end{align}
We restrict ourselves at the bottom of the lower polariton dispersion, where the dispersion can be approximated to be a parabola having effective mass $m$. $V_e$ is the external potential for the polaritons, which consists of microcavity pillars \cite{Klembt2018,PhysRevB.102.121302,Dusel2020}. $\lambda$ is the effective lifetime of the polaritons, which can be controlled by adjusting the quality factor of the cavity. $\alpha>0$ is the effective polariton-polariton interaction coefficient, which is typically repulsive in nature. $f$ is the coherent drive. Next, we perform the following transformation in order to transform to the eq.~(\ref{Eq3})
 \begin{align}\label{Eq19}
&X\rightarrow xa,~ Y\rightarrow ya,~V_e\rightarrow V\varepsilon_u, ~T\rightarrow t t_u,~\Omega_p\rightarrow \omega_p/t_u,\nonumber\\
&\lambda\rightarrow 2t_u/\Gamma,~\phi\rightarrow\psi\sqrt{\frac{\varepsilon_u}{\alpha}},~\text{and}~f\rightarrow F\frac{\varepsilon_u^{3/2}}{\sqrt{\alpha}},
\end{align}
where $x,~y,~V,~\Gamma,~\psi,~\text{and}~F$ are the dimensionless quantities used in eq.~(\ref{Eq3}). Here $a$ is the length unit, $\varepsilon_u=\hbar^2/2ma^2$ is the energy unit, $t_u=2ma^2/\hbar$ is the time unit, $\sqrt{\varepsilon_u/\alpha}$ is the wave function unit, and $\varepsilon^{3/2}_u/\sqrt{\alpha}$ is the pump unit. After the above transformation in eq.~(\ref{Eq19}),  eq.~(\ref{Eq18}) can be matched  exactly with eq.~(\ref{Eq3}). 

We set $a=3~\mu$m, which makes the diameter of the {\it main} pillars 3 $\mu$m and that for the {\it auxiliary} pillars 3.4 $\mu$m. The spacing  between  the {\it main} pillars becomes 6.4 $\mu$m. Such a system of micropillars is readily achieved in experiments \cite{Klembt2018,PhysRevB.102.121302,Dusel2020}.
By choosing $m=5\times 10^{-5}m_e$, where $m_e$ is the free electron mass, the energy unit becomes $\varepsilon_u\approx0.085$ meV. Fixing the energy unit also fixes all the energy scales. The topological bandgap becomes around 0.12 meV, the effective potential depth for the polaritons becomes 20 meV, and the lifetime becomes 30 ps. The energy scale can be increased by choosing  smaller sized micropillars and reducing $a$, which increases $\varepsilon_u$. Alternatively, by adjusting the detuning between the exciton and photon branches, the effective mass of the polaritons can be reduced, thereby increasing $\varepsilon_u$.

Another important parameter is the nonlinear interaction constant $\alpha$. The value of $\alpha$ can be controlled by adjusting the exciton fraction of the polaritons. However, measuring $\alpha$ exactly in experiments is difficult and still an ongoing research. The measurable quantity in experiments is the blueshift $\alpha|\phi|^2$. For the above-mentioned parameters the blueshift for the low-intensity state is around 0.03 meV and that for the high-intensity state is around 0.35 meV. Such values of blueshift are routine observation in experiments \cite{Sun2017,PhysRevB.100.035306}.

For the introduction of the gain at the four corners, an optical pump positioned at the corner having much higher energy from the polariton resonance can be used. The pump creates free electron-hole pairs, which relax down and form the excitonic reservoir. The density of the excitons in the excitonic reservoir act as  gain to the polaritons \cite{RevModPhys.85.299}. Alternatively, it is also possible to arrange gain using electrical pumping \cite{Schneider2013,doi:10.1063/1.4979836,doi:10.1021/acsphotonics.1c01605}.

\section*{Conclusion}
To conclude, we have presented a new concept, where the topological phase can be induced as a memory. In particular, we show that a nonlinear system of photonic lossy resonators goes through a topological phase transition under the application of a coherent pulse. The system continues to maintain its topology, although the effect of the pulse disappears. The topological modes can show fair robustness against realistic disorder (see supporting information). This scheme is independent of the dimension and similar effects can be found in 1D lattices (see supporting information). Our scheme is the first example, where onsite Kerr-nonlinearity induces higher-order topology and can be useful in triggering different functionalities of active topological photonic devices.

\subsection*{acknowledgement}
S. M. thanks Dr. Rimi Banerjee for helpful discussions.

\subsection*{Funding Sources}
This research is supported by the Singapore National Research Foundation Competitive Research Program under Grant No. NRF-CRP23-2019-0007, and the Singapore Ministry of Education Academic Research Fund Tier 3 Grant MOE2016-T3-1-006.

\section*{Supporting Information Available}

This material is available free of charge via the internet at http://pubs.acs.org. The supporting information contains some additional calculations and two movies. The additional calculations include the discussion on bistability, effective tight binding model, calculation of the topological invariant, robustness of the scheme, and 1D topological memory. Movie 1 shows the memory effect without the gain at the corner. Movie 2 shows the triggering of the amplification of the corner modes with a pulse.

\bibliography{Maintext}

\end{document}


\title{Supporting information for Topology with memory in nonlinear driven-dissipative photonic lattices}
\author{Subhaskar Mandal}\email[Corresponding author:~]{subhaskar.mandal@ntu.edu.sg}
\author{Gui-Geng Liu}
\author{Baile Zhang}\email[Corresponding author:~]{blzhang@ntu.edu.sg}

\affiliation{Division of Physics and Applied Physics, School of Physical and Mathematical Sciences, Nanyang Technological University, Singapore 637371, Singapore}

\begin{abstract}
In this supporting information we provide some additional calculations and numerical results.
\end{abstract}

\maketitle

\section*{Bistability for a single resonator}
\begin{figure}[htb]
\includegraphics[width=0.9\textwidth]{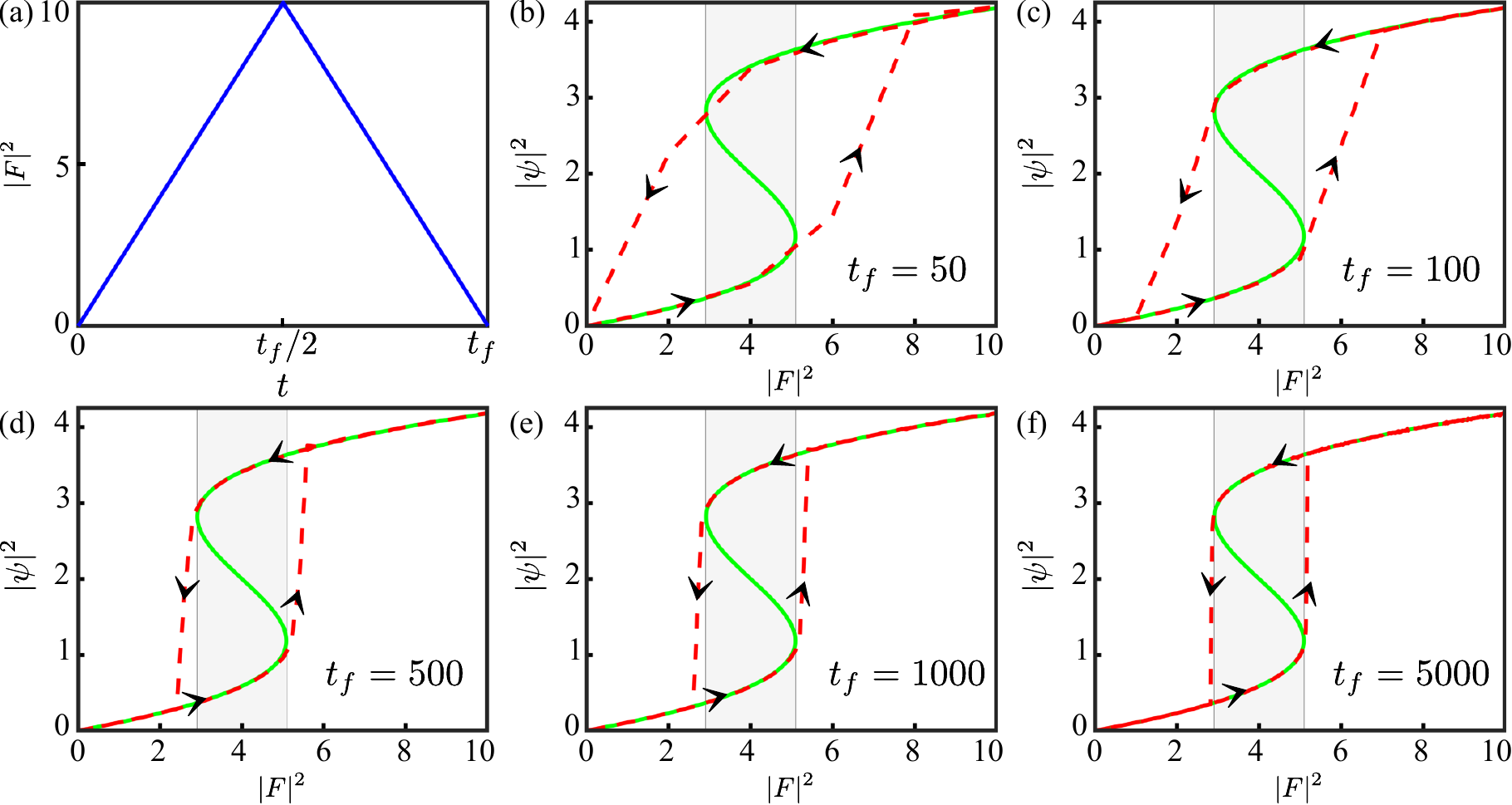}
\caption{(a) The profile of $F$ used for calculating the bistability numerically. $t_f$ is the total calculation time. (b,c,d,e,f) Numerically obtained bistability in red for different values of $t_f$. The green curve corresponds to the analytically calculated bistability. Parameters: All other parameters are kept the same as those in Figure~1b in the main text.}
\label{Fig1}
\end{figure}

Analytical calculation of bistability for a single resonator can be done using eq. (2) in the main text. However, when we move to a more complicated system, such as lattices of coupled resonators, obtaining bistability analytically becomes challenging and one needs to rely on numerical techniques. To recall, the nonlinear schr\"odinger equation (NLSE) is expressed as 
\begin{align}\label{SEq1}
i\frac{\partial \psi}{\partial t}=\left(\omega_0-i\Gamma\right)\psi+|\psi|^2\psi+F(t)\exp(-i\omega_p t).
\end{align}
It should be noted that here $F$ is a function of time $t$. In order to obtain bistability, we need to first slowly increase followed by a slow decrease in $F$ with time. In Figure~\ref{Fig1}a the profile of $|F|^2$ is shown, which has a linear form. We note that the exact profile of $F$ is not important as long as it is first increasing and then decreasing. However, one needs to be careful with the rate at which $F$ is changed. If $F$ is changed too quickly then the system does not get enough time to reach steady-state and in that case, the bistability curve can be inaccurate. In Figure.~\ref{Fig1}b,c,d,e,f the bistability curve for a single resonator is shown for different calculation time $t_f$. For a sufficiently large $t_f$, the numerically calculated bistability matches exactly with  the analytical one.

\begin{figure}[htb]
\includegraphics[width=0.8\textwidth]{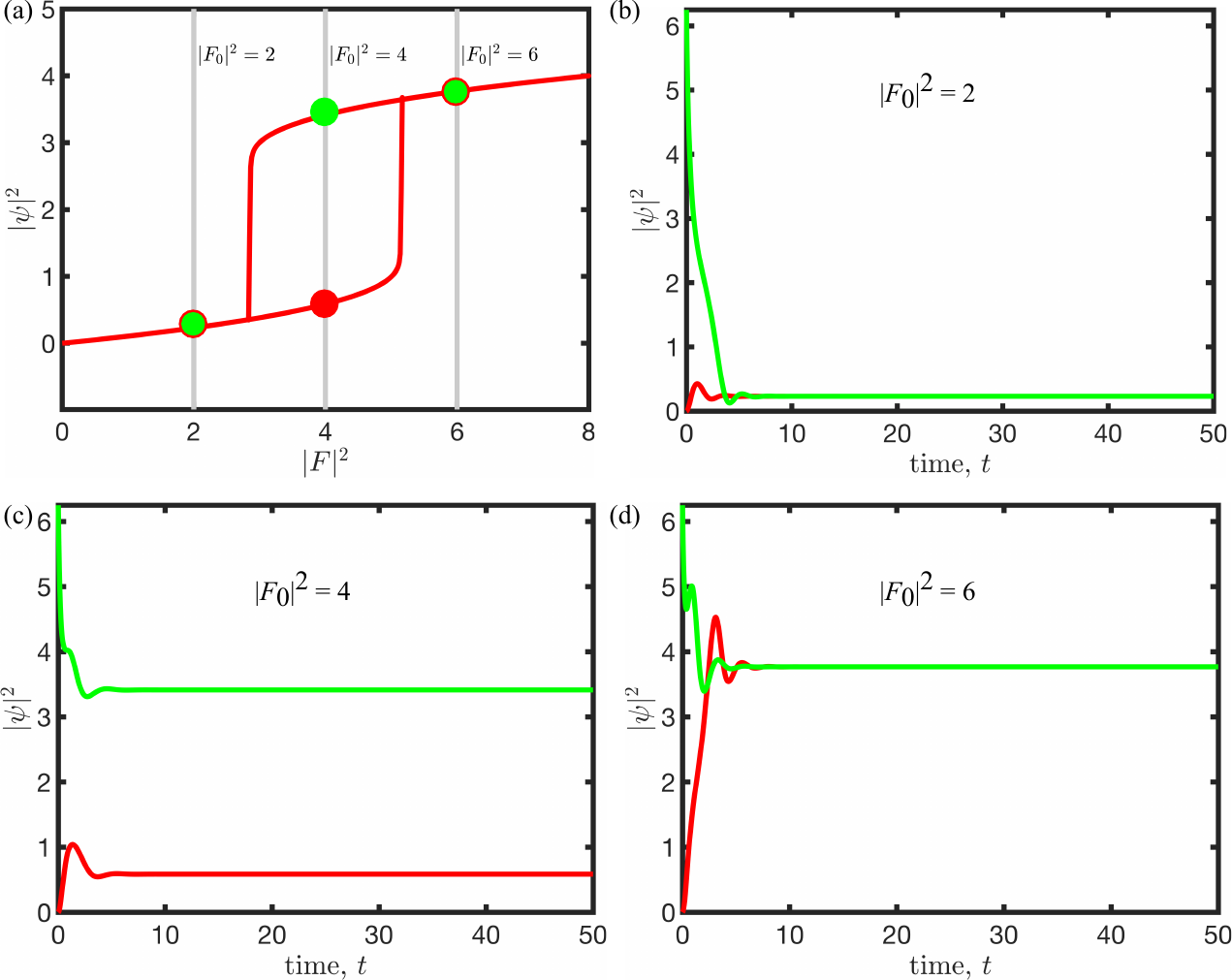}
\caption{(a) The same bistability curve as in Figure~\ref{Fig1}f. The red and green dots correspond to the steady state values taking different initial conditions as shown in (b,c,d). (b,c,d) Time dynamics taking different initial conditions for different $F_0$ values. Different steady states corresponding to different initial condition shows bistable behavior in (c).}
\label{Fig1_2}
\end{figure}

\begin{figure}[htb]
\includegraphics[width=0.5\textwidth]{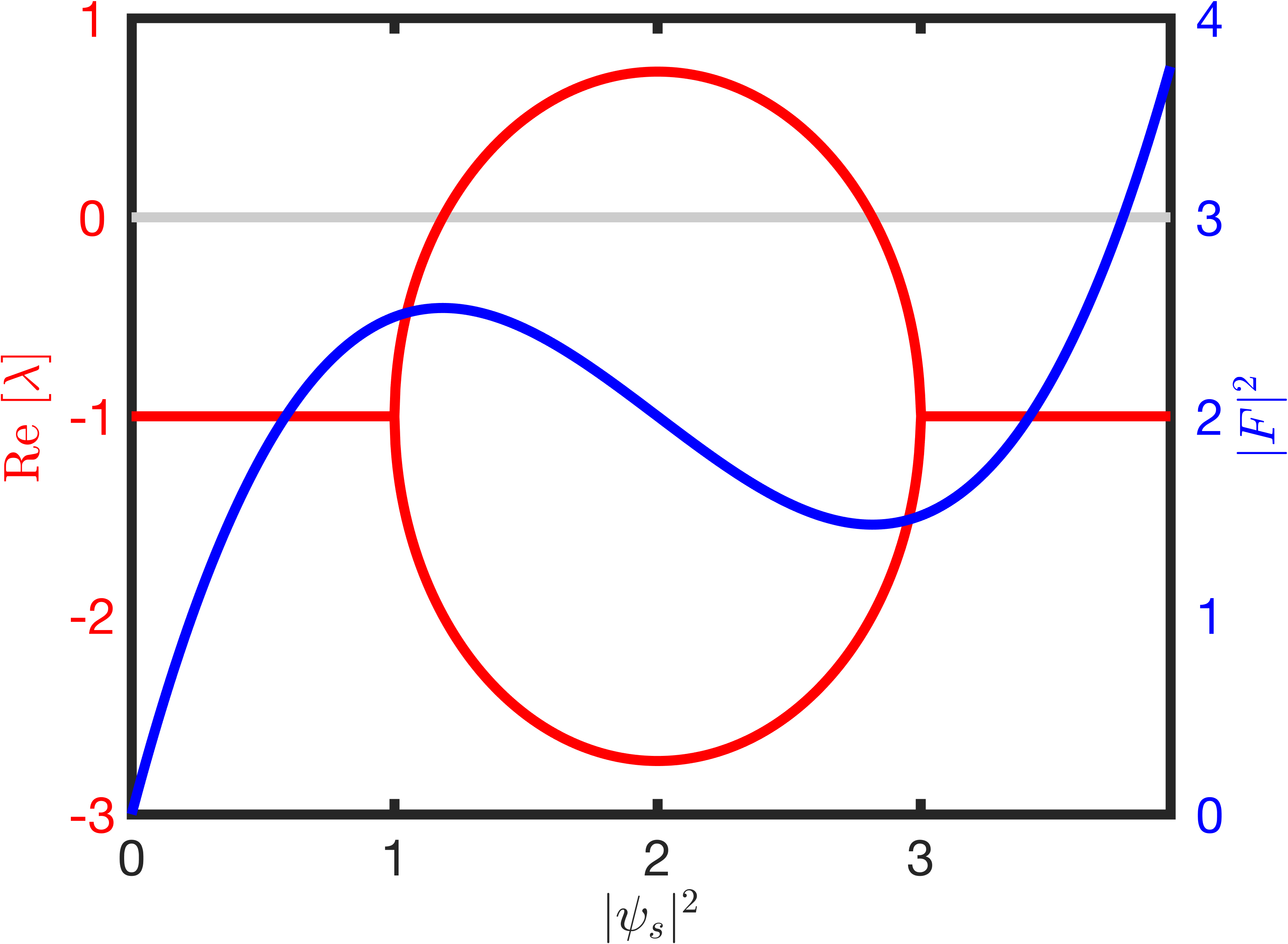}
\caption{The real part of the eigenvalues $\lambda$ of the Jacobian matrix ($\mathcal{J}$) in red and the bistability curve in blue.}
\label{Fig1_3}
\end{figure}

Alternatively, one can perform the time dynamics for each $F$ independently corresponding to two different initial conditions. The steady state values will reproduce the bistability curve. Outside the bistable regime, the steady-state solution does not depend on the initial condition. However, within the bistable regime, large and smaller initial conditions will lead to higher and lower intensity states, respectively. In Figure~\ref{Fig1_2} we have shown such a behavior taking two different initial conditions. It can be seen that when $|F_0|^2$ is outside the bistability region, both initial conditions leads to same steady state. However, when $|F_0|^2$ is inside the bistability region, different initial conditions leads to low and high intensity states. It should be noted that, in the time dynamics, for each value of $F_0$  the steady state is reached for t<10. This is why $t_f=5000(>>10)$ is enough to calculate the accurate bistability curve shown in Figure S1, which at each time step effectively produces the steady state of the system.

While performing the numerical calculations the middle branch of the bistability curve does not appear. This can be explained using First Lyapunov Criterion of stability. To do so, we add some fluctuations $\delta\psi$ on top of  $\psi$. By substituting $\psi\rightarrow(\psi+\delta\psi)\exp(-i\omega_p t)$ in eq.~(1) of the main text we can obtain the following equations
\begin{align}
\frac{\partial\psi}{\partial t}&=-i(\Delta-i\Gamma)\psi-i|\psi|^2\psi-iF,\label{EqS2}\\
\frac{\partial(\delta\psi)}{\partial t}&=-i(\Delta-i\Gamma)\delta\psi-2i|\psi|^2\delta\psi-i\psi^2\delta\psi^*,\label{EqS3}\\
\frac{\partial(\delta\psi^*)}{\partial t}&=i(\Delta+i\Gamma)\delta\psi^*+2i|\psi|^2\delta\psi^*+i{\psi^2}^*\delta\psi\label{EqS4}.
\end{align}

Here $\Delta=\omega_0-\omega_p$, * corresponds to complex conjugation. eq.~(\ref{EqS4}) is obtained by complex conjugation of the eq.~(\ref{EqS3}). The higher-order terms of $\delta\psi$ are ignored. The first equation provides the steady state and the last two equations provide the Jacobian of the form
\begin{align}\label{EqS5}
\mathcal{J}=
\begin{bmatrix}
-i(\Delta-i\Gamma)-2i|\psi|^2&-i\psi^2\\
i{\psi^2}^*&i(\Delta+i\Gamma)+2i|\psi|^2
\end{bmatrix}.
\end{align}
The real part of the eigenvalues $\lambda$ of the Jacobian matrix $\mathcal{J}$ is plotted in red in Figure~\ref{Fig1_3} as a function of the steady-state intensity. The bistability behavior is shown in blue for reference. As it can be seen Re$[\lambda]>0$ only for the middle branch of the bistability curve, indicating that they are not asymptotically stable according to the First Lyapunov Criterion, which is consistent with the numerical calculations.

\section*{Bistability in a lattice}
\begin{figure}[htb]
\includegraphics[width=0.9\textwidth]{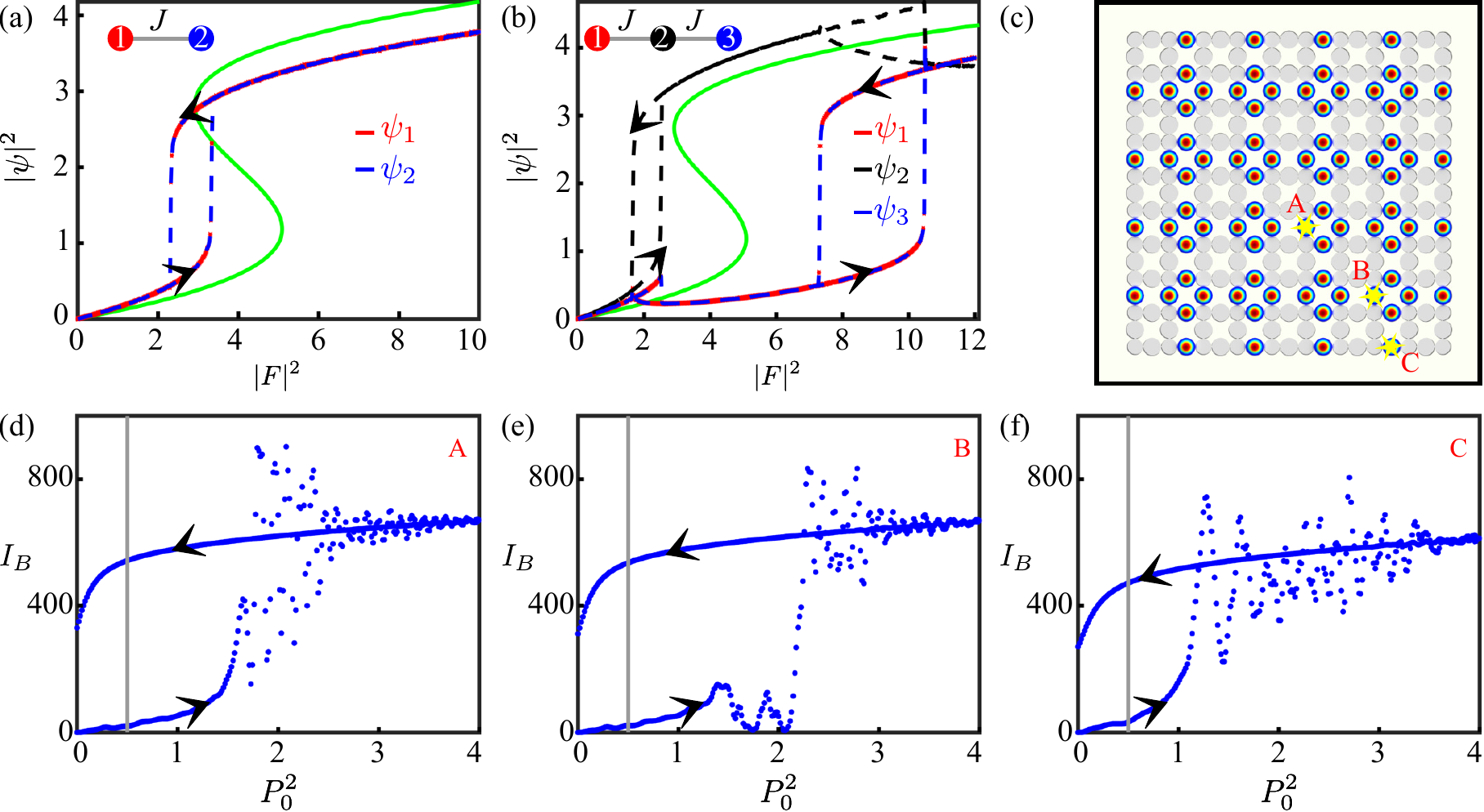}
\caption{(a) Bistability curves for two coupled resonators subjected to a coherent drive (red and blue). (b) Bistability curves for three coupled resonators subjected to a coherent drive (red, black, and blue). The green curve in (a,b) correspond to an isolated resonator case. (c) The system used in the main text, where the yellow stars represent the sites corresponding to which bistability curves are plotted in (d,e,f). To capture the intensity corresponding to a particular site a Gaussian placed on that site is multiplied by the total intensity of the system. Parameters: (a,b) $J=0.5$. All other parameters are kept the same as those in Figure~1 in the main text. (c,d,e,f) All the parameters are kept the same as those in Figure~3 in the main text.}
\label{Fig2}
\end{figure}

Defining bistability corresponding to a single uniform drive $F$ for each resonator in a lattice is extremely challenging. To understand this, we first consider two coupled identical resonators in the rotating frame:
\begin{align}\label{SEq2}
i\frac{\partial \psi_1}{\partial t}&=\left(\Delta-i\Gamma\right)\psi_1+|\psi_1|^2\psi_1+F+J\psi_2,\notag\\
i\frac{\partial \psi_2}{\partial t}&=\left(\Delta-i\Gamma\right)\psi_2+|\psi_2|^2\psi_2+F+J\psi_1.
\end{align}
Here $\Delta=(\omega_0-\omega_p)$ and the last term represents the coupling between the two resonators. Due to the presence of $J$ the bistability curve corresponding to each resonator gets modified. In Figure~\ref{Fig2}a the bistability curves for the two coupled resonators are shown. The curves are identical, however, they are significantly  different from an isolated resonator. This can be understood from the fact that due to the presence of $J$, the effective drive acting on each resonator becomes $F_{\text{eff}}=F+J\psi_1=F+J\psi_2$. For the two resonator case $\psi_1=\psi_2$, which results in the same bistability curves for both resonators. However, if more resonators are introduced the bistability curves for all resonators are no longer the same. In Figure~\ref{Fig2}b the bistability curves for the three coupled resonators are shown, where the middle resonator has a completely different bistability curve from the other two. This is understandable as the middle resonator is coupled with both the  two end resonators, however, the end resonators are coupled with the middle resonator only.

There is no way to obtain exactly identical bistability curves for each resonator in a lattice. One can make them approximately identical by making the coupling $J\rightarrow 0$. However, making $J\rightarrow 0$ is not desirable for studying topological phases in a lattice. This is where the {\it auxiliary} pillars play a crucial role in our scheme as they can be weakly coupled without affecting the relatively strong coupling between the {\it main} resonators. The couplings can be estimated from the flatness of the {\it auxiliary} and {\it main} pillar bands in Figure~4a of the main text. The bistability curves defined on the different {\it auxiliary} pillars that are used in the main text are shown in Figure.~\ref{Fig2}c,d,e,f. The curves have a common bistability region with similar intensities at the chosen $F$. The dissimilarity in the bistability curves results in slightly different intensities in the low and high-intensity states of the {\it auxiliary} resonators, which act as a disorder in the system. However, being topological in nature the topological corner modes remain unaffected.    

\begin{figure}[t]
\includegraphics[width=\textwidth]{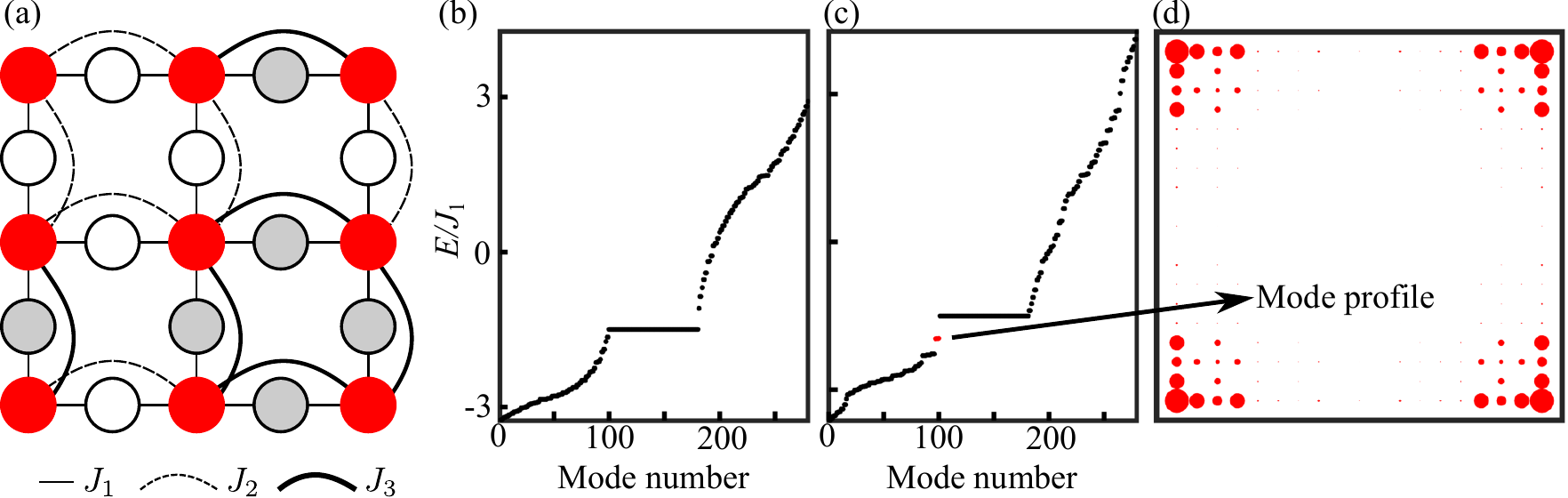}
\caption{(a) Schematic of the system in the tight-binding limit. (b,c) Energy as a function of mode numbers for the case $J_2=J_3$ and $J_3>J_2$, respectively. (d) Spatial profile of a topological corner mode. Parameters: (b) $J_2=J_3=0.3J_1$ (c) $J_2=0.3J_1$ and $J_3=1.2J_1$. The onsite energy of the {\it main} resonators is set to zero and that of the {\it auxiliary} resonators is set to $-1.5J_1$}
\label{Fig_TB}
\end{figure}

\section*{Effective tight-binding model}
In this section, we provide an effective tight-binding model for the scheme presented in the main text, where we consider only the necessary terms needed for the existence of the topological corner modes. In Figure~\ref{Fig_TB}a the structure of the lattice and relevant couplings are shown. The {\it main} resonators are shown in red and the {\it auxiliary} resonators are shown in white and gray. The gray resonators are the ones subjected to the coherent drive. The solid black line represents the nearest neighbour coupling $J_1$ between the {\it main} and {\it auxiliary} resonators. The dashed and solid black curves represent the couplings $J_2$ and $J_3$, respectively, between two neighbouring {\it main} resonators connected by  un-pumped and pumped  {\it auxiliary} resonators, respectively. In the linear regime as well before the pulse, where the intensity in the {\it auxiliary}  resonators is low, $J_2=J_3$. However, after the pulse, the increased intensity in the pumped {\it auxiliary} resonators makes  $J_3>J_2$.

The overall system corresponds to  a Lieb lattice, which is well known to exhibit flat bands in the tight-binding limit. In Figure.~\ref{Fig_TB}b,c the flat band can be seen for a system consisting of ten {\it main} resonators along each of the two directions. When $J_2=J_3$ the system does not show topological modes (see Figure~\ref{Fig_TB}b). However, for  $J_3>J_2$ a bulk bandgap having topological corner modes inside opens up (see Figure~\ref{Fig_TB}c). In Figure~\ref{Fig_TB}d the spatial profile of the corner mode is shown. Although this model effectively captures the physics in our system, we stress that the actual system in the main text is much more complicated. For example, the flat band in the continuum model is not observable due to other next-nearest neighbour hopings that are ignored here. The variation of the onsite terms is also ignored for simplicity. The uniform nearest neighbour coupling $J_1$ also gets modified after the application of the pulse.

\section*{Calculation of the Bloch eigenstates}
To the best of our knowledge, it is not easy (if not impossible) to calculate the topological invariant corresponding to a nonlinear system. This is also acknowledged by other authors \cite{NatElectron.1.178.2018}. However, one can assign a relevant topological invariant corresponding to the Bogoliubov Hamiltonian in eq. (6) of the main text. To do so first we need the Bloch wave functions of the fluctuations. Let's rewrite eq. (6) of the main text in equation form:
\begin{align}
\left[-\nabla^2+V(x,y)-i\Gamma-\omega_p+2|\psi_s(x,y)|^2\right]u_n(x,y)+\psi_s^2(x,y)v_n(x,y)&=\omega_nu_n(x,y),\\
\left[\nabla^2-V(x,y)-i\Gamma+\omega_p-2|\psi_s(x,y)|^2\right]v_n(x,y)-{\psi_s^{2^*}(x,y)}u_n(x,y)&=\omega_nv_n(x,y).
\end{align}
These can be written in a compact form as
 
\begin{align}
\left[-\nabla^2+A_1(x,y)\right]u_n(x,y)+A_2(x,y)v_n(x,y)&=\omega_nu_n(x,y),\label{EqS5}\\
\left[\nabla^2+A_3(x,y)\right]v_n(x,y)+A_4(x,y)u_n(x,y)&=\omega_nv_n(x,y),\label{EqS6}
\end{align}
where, 
\begin{align}
A_1(x,y)&=V(x,y)-i\Gamma-\omega_p+2|\psi_s(x,y)|^2,\notag\\
A_2(x,y)&=\psi_s^2(x,y),\notag\\
A_3(x,y)&=-V(x,y)-i\Gamma+\omega_p-2|\psi_s(x,y)|^2,\notag\\
A_4(x,y)&=-{\psi_s^{2^*}(x,y)}\notag.
\end{align}

The coherent pump $F$ is chosen in a periodic manner, making the steady state $\psi_s(x,y)$ periodic for an infinite system. Under the assumption that the fluctuations are also going to be periodic, we can apply the Bloch theorem to them:
\begin{align}
u_n(x,y)&=U^{k_x,k_y}_n(x,y)\exp\left[i(k_xx+k_yy)\right],\label{EqS7}\\
v_n(x,y)&=V^{k_x,k_y}_n(x,y)\exp\left[i(k_xx+k_yy)\right],\label{EqS8}
\end{align}
where $U^{k_x,k_y}_n$ and $V^{k_x,k_y}_n$ are the Bloch wave functions having Bloch momentum $(k_x,k_y)$. Using the periodicity we can write down the followings:
\begin{align}
U^{k_x,k_y}_n(x+1,y+1)=U^{k_x,k_y}_n(x,y)&=\sum_{G_x,G_y}\tilde{U}_n(G_x,G_y)\exp\left[i(G_xx+G_yy)\right],\label{EqS9}\\
V^{k_x,k_y}_n(x+1,y+1)=U^{k_x,k_y}_n(x,y)&=\sum_{G_x,G_y}\tilde{V}_n(G_x,G_y)\exp\left[i(G_xx+G_yy)\right],\label{EqS10}\\
A_m(x+1,y+1)=A_m(x,y)&=\sum_{G_x,G_y}\tilde{A}_m(G_x,G_y)\exp\left[i(G_xx+G_yy)\right]\label{EqS11},
\end{align}
where $m=1,2,3,4$ and $(G_x,G_y)$ is the reciprocal lattice vector that can take values $2\eta\pi$, where $\eta=0,\pm1,\pm2,...,\infty$. We have set the periodicity in both  directions as 1. The coefficient $\tilde{A}_m$ can be calculated from the following relation:
\begin{align}
\tilde{A}_m(G_x,G_y)=\int_{\text{unit cell}}A_m(x,y)\exp\left[-i(G_xx+G_yy)\right]~dxdy
\end{align}
Substituting eqs.~(\ref{EqS7}-\ref{EqS11}) into eqs.~(\ref{EqS5}-\ref{EqS6}) we can get the following eigenvalue equation, which can be diagonalized to obtain the Bloch states $U_n$ and $V_n$:
\begin{align}
\left[\left(k_x+G_x\right)^2+\left(k_y+G_y\right)^2\right]\tilde{U}_n(G_x,G_y)&+\sum_{G_1,G_2}\tilde{A}_1(G_x-G_1,G_y-G_2)\tilde{U}_n(G_1,G_2)\notag\\
&+\sum_{G_1,G_2}\tilde{A}_2(G_x-G_1,G_y-G_2)\tilde{V}_n(G_1,G_2)=\omega_n\tilde{U}_n(G_x,G_y)\\
-\left[\left(k_x+G_x\right)^2+\left(k_y+G_y\right)^2\right]\tilde{V}_n(G_x,G_y)&+\sum_{G_1,G_2}\tilde{A}_3(G_x-G_1,G_y-G_2)\tilde{V}_n(G_1,G_2)\notag\\
&+\sum_{G_1,G_2}\tilde{A}_4(G_x-G_1,G_y-G_2)\tilde{U}_n(G_1,G_2)=\omega_n\tilde{V}_n(G_x,G_y)
\end{align}
The Bloch eigenstate is given by $\Psi^{n}_{k_x,k_y}=[U^{k_x,k_y}_n,V^{k_x,k_y}_n]^T$.

\section*{Robustness}
\begin{figure}[t]
\includegraphics[width=0.8\textwidth]{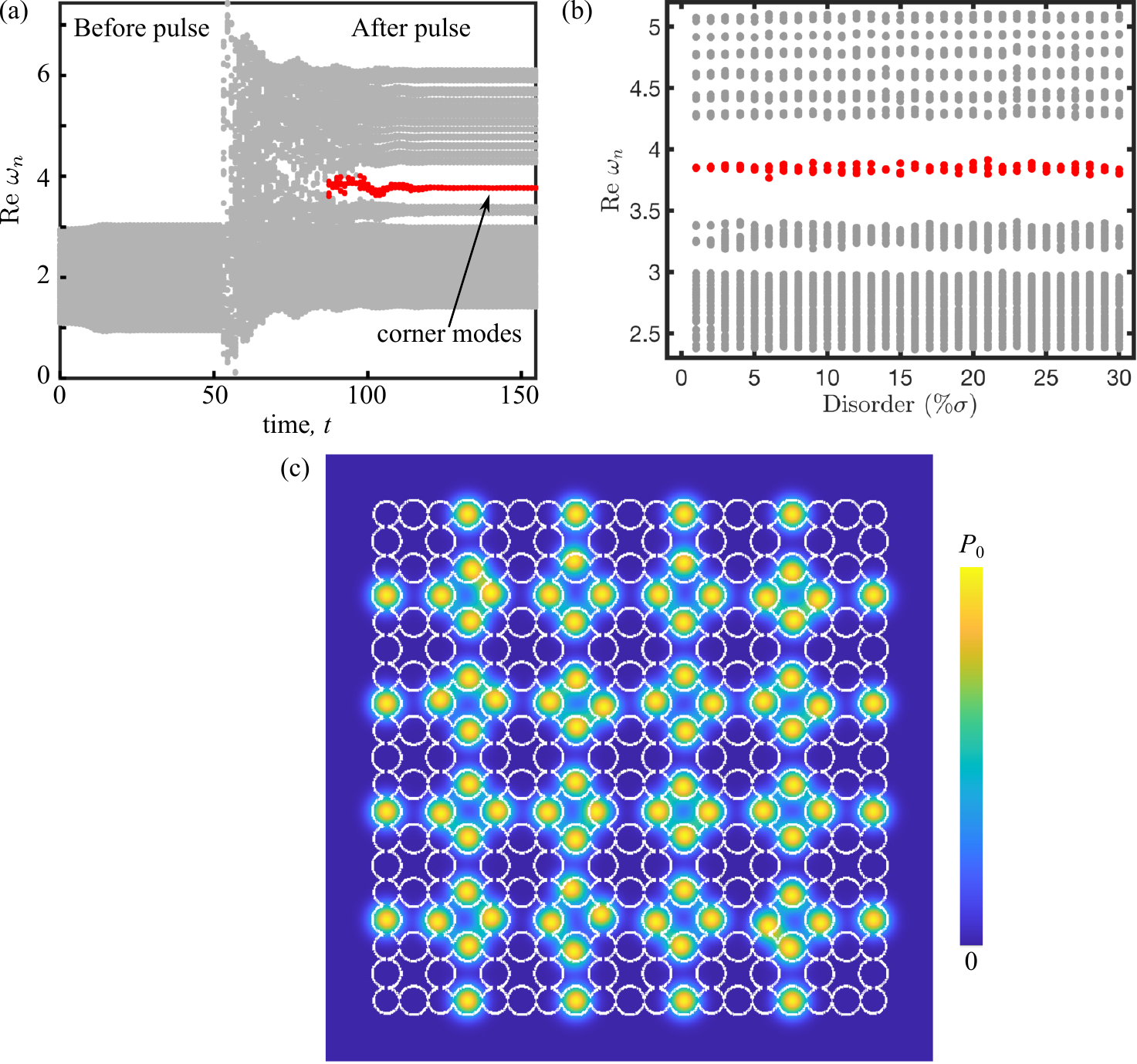}
\caption{(a) Bogoliubov spectrum at each time step in presence of the intensity fluctuation of the coherent drive. The corner modes remain unhampered in presence of the $1\%$ intensity fluctuation of $F$. (b) Bogoliubov spectrum as a function of disorder strength in the pump spot positions. Topological corner modes are shown in red. $\sigma$ is the width of the pump spots. (c) Pump profile with 30\% disorder.}
\label{Fig_Disorder}
\end{figure}

To show the robustness of the topological corner modes as well as the memory effect, we add a time dependent noise $\mathcal{F}(x,y,t)$ at the right hand side of of eq.~(3) in the main text. $\mathcal{F}$ has the same spatial profile as that of $F$ (eq.~(4) in the main text), however the amplitude of the each pump spot decided by the random numbers $P_n$ 
\begin{align}\label{Eq4}
\mathcal{F}(x,y)=\sum_{X_n,Y_n}P_n\exp \left[-\frac{(x-X_n)^2+(y-Y_n)^2}{2\sigma^2}\right].
\end{align}
 We fix $P_n$ such that it corresponds to the $1\%$ fluctuation in the intensity of $F$. We note that it is possible to obtain the intensity of the coherent drive with only $0.02\%$ fluctuations in experiments \cite{PRB.92.165303.2015}.

Next, we solve eq.~(3) in the main text and obtain the Bogoliubov spectrum at each time step, which is shown in Figure~\ref{Fig_Disorder}a. After the application of the pulse around $t=50$, the system switches to the upper intensity state and the corner modes appear, which remain unhampered in presence of the considered intensity fluctuation of the coherent drive. 

Another kind of disorder could arise due to the misalignment of the pump spots. To check the robustness against such kind of disorders, we deliberately add random numbers in the position of the pump spots in the bulk with normal distribution and obtain the steady states followed by the fluctuation spectrum. Figure~\ref{Fig_Disorder}b shows that the topological corner modes (shown in red) remain unhampered with 30\% disorder strength. Here we choose the width of the pump spots  as a disorder unit.  Figure~\ref{Fig_Disorder}. An example of a pump profile with 30\% disorder strength is shown in Figure~\ref{Fig_Disorder}c .

Modern liquid crystal based spatial light modulators can achieve very high accuracy with less than 1\% error. Furthermore, it is also possible to make the steady-state intensity corresponding to  each pump spot uniform by adjusting the pump profile iteratively using the feed-back method based on well known Gerchberg–Saxton algorithm \cite{Gerchberg_Saxton_1972}. Such a method is used to obtain more than 100 tightly-localized states with uniform intensity arranged in square and honeycomb lattices in experiments \cite{Topfer_2021}.

\section*{1D topological memory}

\begin{figure}[t]
\includegraphics[width=\textwidth]{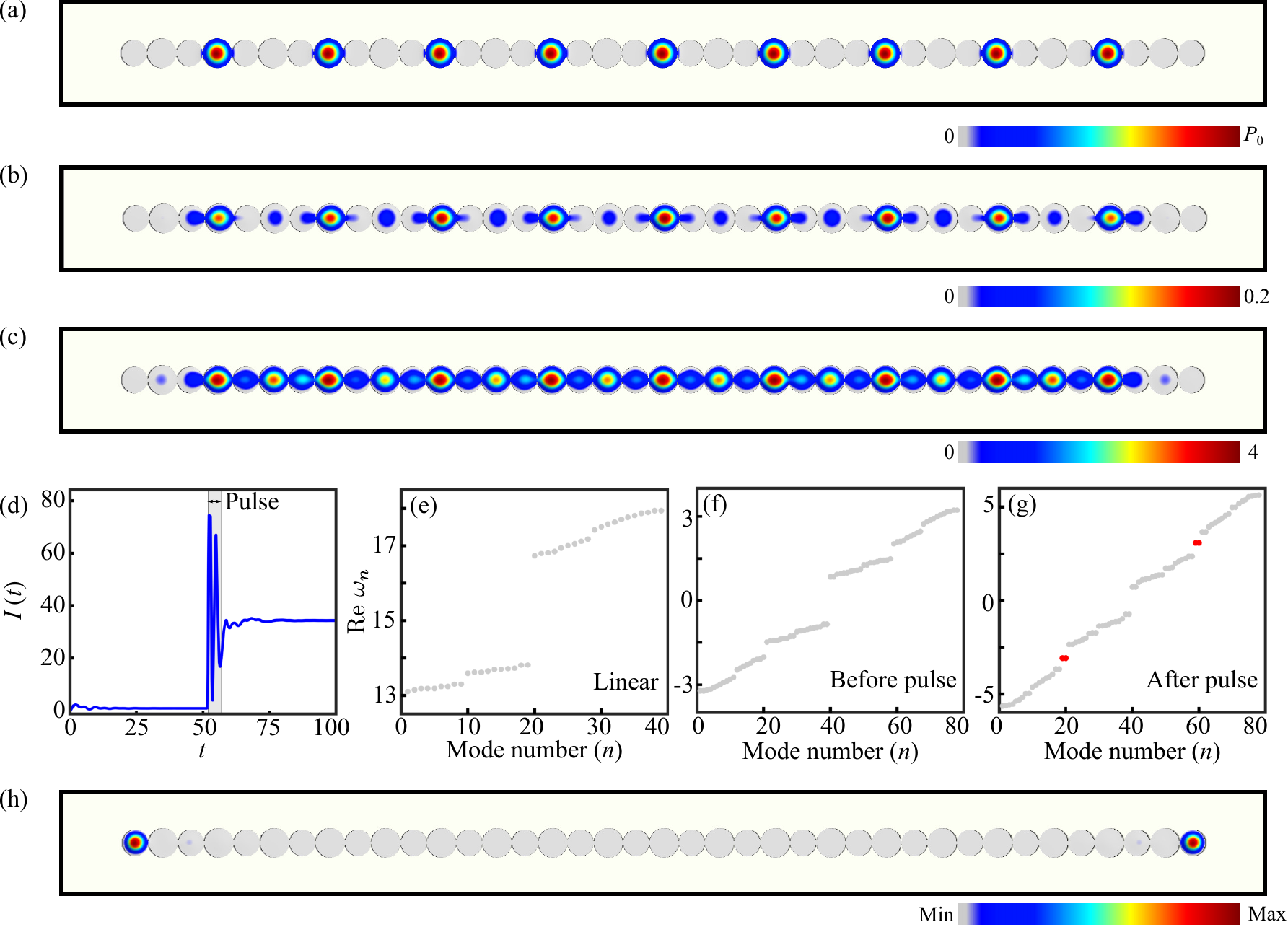}
\caption{(a) The spatial profile of the coherent pump $F$. (b,c) Steady states of the system before and after the coherent pulse $F_p$, respectively. (d) The intensity of the whole chain $I(t)$ as a function of time, which shows the bistable behavior. (e) Real eigenfrequencies of the linear system. (f,g) Real eigenfrequencies of the fluctuations before and after the pulse, respectively. Red dots in (g) correspond to the topological edge modes. (h) The spatial profile of the topological edge mode ($n=60$) induced by the pulse. Parameters: $V_0=236$, $L=2.13$, $d_m=1$, $d_a=1.13$, $\Gamma=0.13$, $P_0=\sqrt{0.4}$, $\omega_p=14.74$, $\sigma=0.3$, $\tau=2.6$, $t_0=50$.}
\label{Fig6}
\end{figure}

Here we show that our scheme is independent of the dimension and demonstrate a 1D topological memory, where the resonators are arranged in a 1D chain. We choose the spatial profile of $F$ such that every alternate {\it auxiliary} resonator is subjected to it (see Figure~\ref{Fig6}a). We choose the pump such that the pumped {\it auxiliary} resonators are in the bistable regime. In Figure~\ref{Fig6}b,c the steady-state before and after the pulse are shown, respectively. Figure~\ref{Fig6}d shows the total intensity of the system $I$ as a function of time, where the pulse enhances the intensity showing 1D memory.

Next, we move on to the topology associated with the steady states. First, we plot the linear spectrum of the system by neglecting the nonlinear and pumping terms. The lower-band is the {\it auxiliary} resonator band, whereas the upper band is the {\it main} resonator band. The linear spectrum shown in Figure~\ref{Fig6}e is topologically trivial and does not show topological edge modes. Next, we apply Bogoliubov theory on top of the steady states shown in Figure~\ref{Fig6}b,c and plot the fluctuation spectrum in Figure~\ref{Fig6}f,g before and after the pulse is applied, respectively. Before the application of the pulse, the spectrum is also trivial. However, after the pulse is applied, the increased intensity induces topological edge modes in the bulk bandgap as shown in red in Figure~\ref{Fig6}g. It should be noted that due to the particle-hole symmetry the spectrum is symmetric with respect to the zero frequency line. In Figure~\ref{Fig6}h the spatial profile of one of the edge modes is shown.

\section{Movies}
Movie1 shows the full dynamics of the system before and after the pulse is applied. The snapshot of the steady-states from this movie is shown in Figure~3 of the main text.

 Movie2 shows the dynamics of the system with gains at the four corners. The steady state before the pulse is applied is the same as the one in Movie1. However, after the application of the pulse, the corner states get amplified. The snapshot of the steady-states from the movie is shown in Figure~5 of the main text.